\documentclass[10pt,aps,pra,twocolumn,showpacs,superscriptaddress]{revtex4-1}
\usepackage[english]{babel} 	
\usepackage[usenames, dvipsnames]{color} 
\usepackage{graphicx} 
\usepackage{bm} 
\usepackage{amsmath} 
\usepackage{amsthm} 
\usepackage{amssymb} 
\usepackage{times} 
\usepackage[pdftex,colorlinks=true,urlcolor= blue,linkcolor=Blue,citecolor=RedViolet]{hyperref}
\usepackage{multirow}
\newcommand{\id}{{\sf 1 \hspace{-0.3ex} \rule{0.1ex}{1.52ex}\rule[-.01ex]{0.3ex}{0.1ex}}} 
\begin{document}
\title{Nonequilibrium properties of trapped ions under sudden application of a laser}
\author{A. A. Cifuentes}
\affiliation{Centro de Ci\^encias Naturais e Humanas, Universidade Federal do ABC, 
             09210-170, Santo Andr\'e, S\~ao Paulo, Brazil}
\author{F. Nicacio}
\affiliation{Centro de Ci\^encias Naturais e Humanas, Universidade Federal do ABC, 
             09210-170, Santo Andr\'e, S\~ao Paulo, Brazil}
\author{M. Paternostro}
\affiliation{Centre for Theoretical Atomic, Molecular, and Optical Physics, 
             School of Mathematics and Physics, Queens University, 
             Belfast BT7 1NN, United Kingdom}
\author{F. L. Semi\~ao}
\affiliation{Centro de Ci\^encias Naturais e Humanas, Universidade Federal do ABC, 
             09210-170, Santo Andr\'e, S\~ao Paulo, Brazil}
\begin{abstract}
Coherent quantum-state manipulation of trapped ions using classical laser fields is a 
trademark of modern quantum technologies. 
In this work, we study aspects of work statistics and irreversibility in a single trapped 
ion due to sudden interaction with the impinging laser. 
This is clearly an out-of-equilibrium process where work is performed through 
illumination of an ion by the laser. 
Starting with the explicit evaluation of the first moments of the work distribution, 
we proceed to a careful analysis of irreversibility as quantified by 
the nonequilibrium lag.
The treatment employed here is not restricted to the Lamb-Dicke limit, 
which allows us to investigate the interplay between nonlinearities and irreversibility. 
We show that in these multiquantum or sideband regimes, 
variation of the Lamb-Dicke parameter causes a non-monotonic behavior 
of the irreversibility indicator. Counterintuitively, we find a working point 
where nonlinearity helps reversibility, making the sudden quench of the Hamiltonian 
closer to what would have been obtained quasistatically and isothermally. 
\end{abstract}
\maketitle

\section{Introduction} 
Quantum control is key to quantum technologies \cite{rabitz2009}. 
Trapping of neutral 
or charged particles, assisted by cooling techniques to bring them to ultracold 
temperatures, form a mature experimental platform to the development of quantum control.  
In particular, laser-manipulated trapped ions are now one of the most 
developed settings for experimental investigation of quantum effects and the 
implementation of basic building blocks needed for quantum storage, 
communication and processing of information \cite{leibfried,blatt,meekhof1996,roos}. 
Along with the development of quantum technologies, where one is usually interested in 
situations far from thermal equilibrium to fully harness the power of quantum coherences, 
there has been an increasing interest in  nonequilibrium thermodynamics of quantum 
systems, sometimes referred to as Quantum Thermodynamics (QT). 
Properly setting the limits in the extraction of useful work and the related problem of 
entropy production in  nonequilibrium processes in quantum systems  lie at the core 
of QT \cite{plastina, various1, campisi2011}. 
Despite being a relatively new subject, QT is a field which grows steadily and rapidly 
\cite{carlisle,jarzinsky,crooks,tasaki,various2,various3,various4,various5}. 
QT brings the old realm of classical thermodynamics to a new perspective where 
quantum correlations and quantum coherence might play an important role.

Taking all this into account, it seems crucial to investigate the interaction of 
laser fields and trapped ions from a statistical  nonequilibrium perspective, 
trying to uncover new aspects and physically relevant information previously untouched 
by standard approaches to this subject. 
The scenario is particularly rich given the multitude of different energy structures 
and transitions involving electronic and vibrational degrees of freedom accessible in 
this system through laser interaction \cite{orszag,leibfried}. 
One of the motivations for the present work is the possibility of studying 
thermodynamical implications of nonequivalent physical regimes, some of them driven 
by strongly nonlinear Hamiltonians \cite{orszag,leibfried}. 
The interplay between the physics of trapped ions and QT has been explored previously, 
for instance, in the context of ion-based thermo-engines \cite{abah} 
and the verification of fluctuation relations \cite{varios6}. 
However, our work departs from those in both context and methodology.
In particular,  we bring
the variety of physical regimes available in the trapped ion system to the light of 
 nonequilibrium QT. 
This includes carrier and sideband regimes, where single- or multiple-quanta transitions 
and energy dependent couplings manifest 
according to the laser frequency \cite{leibfried}.
We will be particularly interested in a study that addresses thermodynamic irreversibility 
as quantified by the irreversible lag \cite{daffner} produced by a finite-time transformation 
experienced by a trapped ion. 

This paper is organized as follows. 
In Section~\ref{qt}, we briefly review the relevant concepts of nonequilibrium 
QT needed to the subsequent developments. 
This includes a discussion of the microscopic view of work and the basic physics 
upon which the nonequilibrium lag is built. Sec.~\ref{iiwaef} is dedicated 
to a brief review of the laser-manipulated trapped ion system. 
Our results are presented in Section~\ref{tfti}, 
where we discuss work and irreversibility when the trapped ion is suddenly illuminated 
by a classical laser field. In Section~\ref{conc}, we conclude our findings and, 
in the Appendix, we present expressions for  the eigenvalues and eigenvectors of 
the system Hamiltonian for an arbitrary sideband.
%
\section{Work and Irreversibility}\label{qt}      
A open system can exchange energy and/or particles with the environment or 
an external agent. In this context, work is energy which is 
transferred/extracted to/from 
the system through application of arbitrary generalized forces \cite{balian}. 
From a microscopic perspective, work is then necessarily accompanied by a modification 
of the system Hamiltonian (energy levels). This is to be distinguished from heat which
is energy exchanged with the environment through their mutual weak (infinitesimal) 
interaction. Consequently, the effect of heat is not a significant modification of 
the energy levels but a redistribution of their population. 
This too corresponds to a variation of internal energy just like work. 

In general, a nonisolated system may suffer both processes. 
However, in what follows, we will be interested in the scenario where work is 
performed without heat exchange. 
This can be physically achieved, for instance, when the system is thermally isolated 
or the work protocol is performed in a time interval which is orders of magnitude 
shorter than the thermalization time. This is precisely the case of the idealized 
process of sudden Hamiltonian quench which corresponds to the instantaneous change 
of the system Hamiltonian from 
$\hat{\mathcal H}(\lambda_i)$ to $\hat{\mathcal H}(\lambda_f)$. 
In these expressions, $\lambda$ is a macroscopic variable in the system Hamiltonian 
usually called {\it work parameter} in the context of  nonequilibrium thermodynamics.
Work $W$ is a random variable 
encompassing both thermal and quantum fluctuations. In the case of a sudden quench, 
the statistical moments of the work distribution read \cite{fusco2014} 
\begin{equation}                                                                         \label{mom}
\left\langle W^{n}\right\rangle  =  
\sum_{k=0}^{n} \left(-1\right)^{k}
     \begin{pmatrix}n \\ k \end{pmatrix}
     \text{Tr} \left[  \hat{\mathcal{H}}_{\!f}^{\left(n-k\right)} 
                       \hat{\mathcal{H}}_{i}^{k} \, 
                       \hat{\rho}_{i} \right],
\end{equation}
with $n$ integer and 
$\hat{\mathcal{H}}_{\!j} := \hat{\mathcal{H}}(\lambda_j)$ for $j = i,f$. 
More details about the statistical meaning of work can be found in \cite{WGF} and references therein. 

One of the most important results of  nonequilibrium statistical mechanics is the
Jarzinsky equality \cite{jarzinsky},
from which one can directly obtain a fundamental inequality involving average  
work $\left\langle W \right\rangle$ and Helmholtz free energy $F$ 
\begin{eqnarray}                                                                         \label{ji}
\left\langle W \right\rangle \ge \Delta F,
\end{eqnarray} 
where $\Delta F \equiv F(\lambda_f, \beta) - F(\lambda_i,\beta) $ 
is the difference between the free energies of the system. 
Explicitly, 
\begin{equation}                                                                         \label{hfe}
F(\lambda_j, \beta) = - \frac{1}{\beta} {\rm ln}\, \mathcal Z(\lambda_j), 
\,\,\, 
{\mathcal Z} (\lambda_j) \equiv 
                         {\rm Tr} \, {\rm e}^{-\beta \hat {\mathcal H}(\lambda_j) },
\end{equation}
with $j=i,f$. 
The equality in Eq.~(\ref{ji}) is only achieved by an isothermal quasistatic process, which is {\it reversible}  \cite{WGF}.

The indicator of irreversibility used in this work can then be defined 
considering what has just been exposed. Based on Eq.~(\ref{ji}), 
one defines \cite{crooks, daffner}
\begin{equation}                                                                         \label{nl1}
\mathcal L \equiv \beta ( \langle W \rangle - \Delta F),
\end{equation}
as an indicator of irreversibility in the sense that the work protocol 
is reversible only when $\mathcal L=0$. 
What is reversible or irreversible for this indicator 
is the work protocol realized in an initially equilibrated system. 
The idea is that a backwards run of the work protocol after the system starts 
attempting thermal reequilibration will not, in general, bring the system and 
environment to their initial state. The quantity between the parentheses is known as 
irreversible work \cite{crooks}, and $\mathcal L$ is usually 
called ``nonequilibrium lag'' (NL) as it gives an idea of how the system state, 
after the work protocol, lags behind an equilibrium thermal state fixed by the final 
Hamiltonian and inverse temperature $\beta$. 
Remarkably,  it has been shown that the NL is exactly equal 
to the relative entropy between the thermal state used to evaluate 
$F(\lambda_f, \beta)$ and the postwork state \cite{daffner}. 
It is important to remark that the relative entropy is zero for identical states 
and it diverges for orthogonal states \cite{vedral}.

\section{Short Review on Trapped Ions Interacting with            
         Classical Laser Fields} \label{iiwaef}                   
We now present the basic elements needed to work with trapped ions subjected to laser 
fields. 
More information can be found in the many reviews available in the literature, 
e.g., \cite{leibfried}.
Usually, the laser-ion setup is described by a model consisting 
of a two-level system (electronic degrees of freedom) 
coupled to a harmonic oscillator (center of mass motion). 
The latter is the result of electromagnetic confinement achieved by the use of 
trapping technology, e.g., Paul traps \cite{ghosh}, and the electronic-motion coupling 
occurs due to momentum exchange with the laser.   

By considering the center of mass (CM) degree of freedom as an oscillator with 
natural frequency $\nu$, and the two levels  
$\{ | g \rangle$, $ | e \rangle \}$
with an energy separation of $\hbar\omega_0$, 
the system Hamiltonian reads \cite{blockley}
\begin{equation}                                                                         \label{hamtot}
\hat{\mathcal{H}} =  \hat{\mathcal{H}}_0 + \hat{\mathcal{H}}_{\rm I},
\end{equation}
with
\begin{equation}                                                                         \label{hamfree}
\hat{\mathcal{H}}_0 = \hbar\nu\hat{a}^{\dagger}\hat{a} + 
                       \frac{ \hbar \omega_0 }{2} \hat{\sigma}_{z},
\end{equation}
and
\begin{equation}                                                                         \label{Hamint}
\!\!\!\!\hat{\mathcal{H}}_{\rm I} = \frac{\hbar \Omega}{2} 
                         \left[ \hat{\sigma}_{+} \, 
                                \text{e}^{ i \eta ( \hat{a} + \hat{a}^{\dagger} ) 
                                         - i \omega_L t } +  
                                \hat{\sigma}_{-} \, 
                                \text{e}^{ - i \eta ( \hat{a} + \hat{a}^{\dagger} ) 
                                         + i \omega_L t }
                         \right],
\end{equation}
where  $\omega_L$ is the laser frequency, $\Omega$ the classical Rabi frequency,
$\hat a$  the annihilation operator for the CM motion, 
$\hat \sigma_z = | e \rangle \! \langle e | - | g \rangle \! \langle g |$, 
$\hat\sigma_+ = \hat\sigma^\dag_- = | e \rangle \! \langle g |  $, 
and $\eta$ the Lamb-Dicke parameter defined as 
\begin{eqnarray}\label{etatrue}
\eta=\frac{\omega_L}{c}\sqrt\frac{\hbar}{2M\nu}\cos\phi,
\end{eqnarray}
with $M$ being the mass of the trapped ion, $c$ the speed of light, 
and $\phi$ the angle between the laser wave vector and the trap axis 
(one dimensional motion).

Depending on the detuning $\omega_0 - \omega_L$, the laser will cause the coupling of 
different vibrational levels with electronic part, each case representing a different 
quantum-optical process \cite{wineland} with its own effective Hamiltonian. 
The procedure to reveal each of those Hamiltonians is very well described 
in the literature, e.g., \cite{leibfried,orszag}. 
Basically, after setting $\omega_0 - \omega_L=\pm m\nu$, with $m=0,1,2,\ldots$, 
one applies a rotating wave approximation (RWA) to Hamiltonian (\ref{hamtot}) in order to obtain 
\begin{equation}                                                                         \label{hamrwa}
\hat{\mathcal{H}}^{(m)}_{\pm} = \hat{\mathcal{H}}_{0} + 
\hbar \left( \text{e}^{-i\omega_{L} t }\hat{\Omega}_{m}^\pm \hat{\sigma}_{+} + 
             \text{e}^{ i\omega_{L} t }\hat{\Omega}_{m}^\mp \hat{\sigma}_{-} \right), 
\end{equation}
where  
\begin{equation}                                                                         \label{Omegaux1}
\hat{\Omega}_{m}^{+} = \hat{\Omega}_{m}^{- \dag} =  
                       \frac{\Omega}{2} \text{e}^{-{\eta^{2}\!}/{2}}
                       \sum_{l=0}^{\infty}\left(i\eta\right)^{2l+m} 
                       \frac{\hat{a}^{\dagger l}\hat{a}^{l}}{l!(l+m)!} \hat a^m. 
\end{equation}
For consistency, one must notice that $\eta$ in Eq.~(\ref{etatrue}), 
besides being  a function of $\phi$ and $\nu$, 
is also a function of $\omega_0$. This is so because
$\omega_L$ is now fixed by the sideband choice (value of $m$). 

The Hamiltonian $\hat{\mathcal{H}}^{(m)}_{+} $ is obtained with  $\omega_0 - \omega_L=
 m \nu $, and
it describes a $m$-phonon process for the vibrational part accompanied with transitions 
in the atomic levels. 
It can be referred to as a $m$-phonon Jaynes-Cummings (JC) model.   
On the other hand, Hamiltonian $\hat{\mathcal{H}}^{(m)}_{-} $ 
is obtained with $\omega_0 - \omega_L= - m \nu $ 
and it can be referred to as a $m$-phonon anti-Jaynes-Cummings (AJC) model.
The case $m=0$ can be studied using either $\hat{\mathcal{H}}^{(m)}_{+} $ 
or $\hat{\mathcal{H}}^{(m)}_{-} $,  
\begin{equation}                                                                        \label{hamct}  
\!\!\hat{\mathcal{H}}^{(0)}=\hat{\mathcal{H}}^{(0)}_{\pm} = 
\hat{\mathcal{H}}_{0} + 
\frac{\hbar}{2} \left(\! \text{e}^{-i\omega_{L} t }\hat\Omega_0^+ \hat{\sigma}_{+} +
                              \text{e}^{ i\omega_{L} t } \hat\Omega_0^-\hat{\sigma}_{-}\! \right),
\end{equation}
and it describes Rabi oscillations between electronic levels, i.e., the carrier transitions \cite{leibfried,orszag}.

For what comes next, it is useful to present now the matrix elements of 
$\hat{\Omega}_{m}^\pm$ in the Fock basis of the CM harmonic motion
\begin{eqnarray}                                                                          \label{Omegme}
\left\langle n\right|\hat{\Omega}_{m}^+\left|n'\right\rangle &=&
\left\langle n'\right|\hat{\Omega}_{m}^{-}\left|n\right\rangle^\ast                       \\ 
&=& \frac{\Omega(i\eta)}{2}^{\!\!^m} 
 \sqrt{ \frac{ n!}{(m+n)!} } \, \text{e}^{-\eta^{2}/2} 
             L_{n}^{m}\!\left(\eta^{2}\right) \delta_{n'\,n+m},                           \nonumber          
\end{eqnarray}
with the associated Laguerre polynomials \cite{gradshteyn}
\begin{equation}
L_{n}^{m}\left( x \right)  = 
\sum_{k=0}^{n}\left(-1\right)^{k}
\frac{(n + m)!}{(m + k)!(n-k)!}
\frac{x^{k}}{k!} .
\end{equation}

As it can be seen from Eq.~(\ref{Omegme}), the quantum Rabi frequencies, 
$\left\langle n\right|\hat{\Omega}_{m}^+\left|n'\right\rangle$, 
have a strong dependence on the Lamb-Dicke parameter $\eta$. 
For small values of $\eta$, they present a quasilinear dependence on $n$, 
typical of $m$-photon Jaynes-Cummings models in the context of 
cavity quantum electrodynamics (cQED) \cite{multi}. 
However, for the ionic system, it is possible to induce considerable nonlinearities in 
$\left\langle n\right|\hat{\Omega}_{m}^+\left|n'\right\rangle$ simply by increasing 
the Lamb-Dicke parameter. The quantum Rabi frequencies become an oscillating function 
of $n$ due to the presence of the Laguerre polynomials in Eq.~(\ref{Omegme}). 
These oscillations can strongly influence the system dynamics as thoroughly 
studied in \cite{gregorio}. 

\section{Results} \label{tfti}                         
The work protocol we have in mind is now explained. 
First, the work parameter $\lambda_t$ here has to do with the application of the laser 
on the ion. More specifically, we take $\lambda_i=0$ and $\lambda_f=\Omega$ in a sudden 
quench of the system Hamiltonian. This means an abrupt change from 
\begin{eqnarray}\label{h0}
\hat{\mathcal H}(\lambda_i)= \hbar\nu\hat{a}^{\dagger}\hat{a} + 
                       \frac{ \hbar \omega_0 }{2} \hat{\sigma}_{z}
\end{eqnarray}
to
\begin{equation}\label{hf}
\!\!\!\hat{\mathcal H}(\lambda_f) =
                       \hat{\mathcal H}(\lambda_i) + \frac{\hbar \Omega}{2} 
                         \left[ \hat{\sigma}_{+} \, 
                                \text{e}^{ i \eta ( \hat{a} + \hat{a}^{\dagger} ) 
                                         } +  
                                \hat{\sigma}_{-} \, 
                                \text{e}^{ - i \eta ( \hat{a} + \hat{a}^{\dagger} ) 
                                          }
                         \right]\!,
\end{equation}
or, if we want to explore the sidebands, an abrupt change to 
\begin{eqnarray}\label{hs}
\hat{\mathcal H}(\lambda_f) = \hat{\mathcal H}(\lambda_i) + 
\hbar ( \hat{\Omega}_{m}^\pm \hat{\sigma}_{+} + 
        \hat{\Omega}_{m}^\mp \hat{\sigma}_{-} ).
\end{eqnarray}
The above Hamiltonians, Eq.~(\ref{hf}) and Eq.~(\ref{hs}), correspond to the sudden application 
of the laser field, i.e., the result of taking the limit of $t\rightarrow 0$ 
in Eq.~(\ref{Hamint}) and in Eq.~(\ref{hamrwa}), respectively.

It is well known that the Hamiltonian (\ref{hf}) can not be diagonalized exactly, 
so that much of the analytical advances take place with the sideband Hamiltonians 
in Eq.~(\ref{hs}). 
It is important to remark that Eq.~(\ref{hs}) indeed describes quite well the system when 
$\omega_0 - \omega_L=\pm m\nu$ and $\Omega$ is moderately weak, 
which are conditions easily implemented in the laboratories \cite{meekhof1996,roos}. 
Before the interaction with the laser, the trapped ion is found to be in thermal 
equilibrium with the environment  (at inverse temperature $\beta$). 
This is described by the Gibbs state associated with Hamiltonian Eq.~(\ref{h0}), i.e.,
\begin{equation}                                                                         \label{initstate}
\hat{\rho}_i  =  \frac{\text{e}^{-\beta\hbar\nu\hat{n}}}{(\bar n + 1)} \otimes 
\frac{\text{e}^{-\frac{\beta\hbar\omega_{0}}{2}\hat{\sigma}_{z}}}
                      {2\cosh \frac{\beta\hbar\omega_{0}}{2}},
\end{equation}
where $\hat n = \hat a^\dag \hat a$ is the number operator and 
\begin{equation}                                                                         \label{mocn}
\bar n = {\rm Tr}(\hat n \hat \rho_i) = ({\rm e}^{\beta\hbar\nu} - 1)^{-1}
\end{equation}
is the thermal occupation number of the CM motion. 

In spite of the difficulties found in dealing with the full Hamiltonian Eq.~(\ref{hf}), 
we were able to find the first moments of the work distribution. 
This is already valuable information because to obtain the full distribution we would 
need the whole set of eigenvalues and eigenvectors of Eq.~(\ref{hf}) which are not possible 
to be obtained, except numerically and to a restricted precision giving the complexity of 
the Hamiltonian. We then use Eq.~(\ref{mom}), appropriate to a sudden change, 
to calculate a few first moments of the work distribution and get some 
insight of it. 

The first moment, $n=1$, using Eq.~(\ref{hf}) and Eq.~(\ref{initstate}), turns out to be 
\begin{equation}                                                                         \label{1mom}
\langle W \rangle = 
{\rm Tr}\left[ \hat{\mathcal{H}}_{\rm I} \, \hat \rho_i \right] 
\propto {\rm Tr}[\hat \sigma_{\pm} \, \text{e}^{-\frac{\beta\hbar\omega_{0}}{2}
\hat{\sigma}_{z}}] = 0.
\end{equation}
As for the second, we now find
\begin{equation}
\left\langle W^{2}\right\rangle   =  \hbar^2\Omega^2/4,
\end{equation}
which, interesting enough, depends only on the magnitude of the work parameter $\Omega$
(controlled by laser power) and it is completely independent of the temperature.
Since $\langle W \rangle = 0$, the second moment is also the variance 
of the work distribution. 
The third moment is given by  
\begin{equation}                                                                         \label{3mom}
\left\langle W^{3}\right\rangle  = \frac{\hbar^{3}\Omega^{2}}{4}
                                   \left[\nu\eta^{2} + 
                                         \omega_{0}
                                         \tanh \tfrac{\beta\hbar\omega_{0}}{2}
                                   \right],
\end{equation}
in which appears the dependence on the temperature. From the second and third moments, 
we can determine the skewness of the work distribution 
$\left\langle W^{3}\right\rangle/\left\langle W^{2} \right\rangle^{3/2} $. 
This turns out to be inversely proportional to the magnitude of the work parameter. 
Consequently, the stronger the laser, the more symmetric the distribution is around the 
mean value $\left\langle W\right\rangle = 0$. 
Since $\left\langle W^{3} \right\rangle > 0$, as seen from Eq.~(\ref{3mom}), 
the work distribution is biased towards negative values of work.  
All these facts about the first moments of the work distribution, obtained with 
the full Hamiltonian Eq.~(\ref{hf}), tell us that negative work (internal energy descrease) 
is more likely than the equivalent positive work (internal energy increase) at the very 
first instant of interaction with the laser field.     
Note also that the asymmetry around 
the mean value decreases with the temperature while it increases with $\eta$. 
Finally, according to  Eq.~(\ref{etatrue}), $\left\langle W^{3}\right\rangle$ and 
the skewness are actually independent 
of the trap frequency $\nu$.

Now we turn our attention to the sideband Hamiltonians in Eq.~(\ref{hs}) 
and to the irreversibility of the work protocol consisting of the sudden quench of system 
Hamiltonian due to laser interaction. 
As said before, these effective Hamiltonians are obtained from the full 
Hamiltonian Eq.~(\ref{hf}) 
by setting resonance $\omega_{0} = \omega_{L} \pm m \nu$ and performing a 
rotating wave approximation. We will see that a thermodynamic analysis is able 
to reveal the different aspects of the optical processes raised by the selection 
of distinct sidebands.

We proceed to apply the NL in Eq.~(\ref{nl1}) to reveal the irreversibility of the work 
protocol. 
Just like what happened with the full Hamiltonian (\ref{hf}), 
the first moment of the work distribution or simply the average work is again null, i.e.,
$\langle W \rangle = 0$. For this reason, the NL in Eq.~(\ref{nl1}) for the sudden quench 
of the sideband Hamiltonian in Eq.~(\ref{hs}) reads  
\begin{equation}\label{nlfinal}
\mathcal L =  {\rm ln}\, \frac{\mathcal Z(\lambda_f)}{\mathcal Z(\lambda_i)},
\end{equation}
with
\begin{equation}\label{zi}
\mathcal Z(\lambda_i) =  2(\bar n + 1) \cosh{ \tfrac{\beta \hbar\omega_0}{2}},
\end{equation}
obtained using Eq.~(\ref{h0}), and 
\begin{equation}\label{zf}
\!\!\mathcal Z_{\pm}(\lambda_f)\!  = \!
\sum_{n = 0}^{\infty} \left[ {\rm e}^{-\beta \mu_{\pm}^{ (n,m) } } \!\! + \!
                             {\rm e}^{-\beta \gamma_{\pm}^{(n,m)}}\right] + 
\sum_{n = 0}^{m-1}           {\rm e}^{-\beta \zeta_{\pm}^{(n,m)}} \!\! 
,
\end{equation}
obtained with Eq.~(\ref{hs}). 
The functions $\mu_{\pm}$, $\gamma_{\pm}$, and $\zeta_{\pm}$ 
are the eigenvalues of the Hamiltonians in (\ref{hs}), and their expressions 
can be found in Eqs.~(\ref{eigval1+}), (\ref{eigval+}), (\ref{eigval1-}), 
and (\ref{eigval-}), which allows one to get
\begin{eqnarray}                                                                         \label{partf1}
\!\!\!\!\!\!\!\mathcal Z_{\pm}(\lambda_f)  &=& 2  
\sum_{n=0}^{\infty} \!{\rm e}^{-\beta \hbar \nu( n + \frac{m}{2})}
     \!  \cosh\!\!\left[ \tfrac{\beta \hbar}{2} 
     \sqrt{{\omega_L}^{2} \! +\! \Omega^{2}\left|f_n^m\right|^{2}} \right]               \nonumber\\ 
&& + \, 
(\bar n + 1 ) (1- {\rm e}^{- \beta \hbar m \nu} )
{\rm e}^{\pm \tfrac{1}{2}\beta\hbar \omega_0 } , 
\end{eqnarray}
with $\omega_L =  \omega_{0} \mp m\nu$, and
\begin{eqnarray}                                                                          \label{auxf1}
\!\!\!\!\!\!f_n^m\!:=\! \frac{2}{\Omega} \!
             \left\langle n\right|\!\hat{\Omega}_{m}^+\!\left|n+m\right\rangle 
          = \!{(i\eta)}^{\!\!^m}\!\! \sqrt{\!\!\tfrac{ n!}{(m+n)!} } \, 
                 \text{e}^{-\frac{\eta}{2}^{\!2}} L_{n}^{m}\!\left(\eta^{2}\right),       
\end{eqnarray}
where we used Eq.~(\ref{Omegme}). 
With the above two partition functions, we can calculate $\mathcal L$ in Eq.~(\ref{nlfinal}). 
Before the presentation of the simulations, we want to make the notation clear emphasizing 
that $\mathcal Z_{+}(\lambda_f)$ refers the JC-type Hamiltonians with 
$\omega_L =  \omega_{0} - m\nu$, while $\mathcal Z_{-}(\lambda_f)$ 
refers to the AJC-type Hamiltonians with $\omega_L =  \omega_{0} + m\nu$. 

Now we carry on to the numerical investigation of the NL. 
For that, it is important to have in mind the reality of the physical 
parameters to be used in the simulations. 
First, the initial thermal occupation numbers $\bar{n}$ 
will be considered relatively small in order to have quantum fluctuations 
playing some role. 
The experiments employ sophisticated and very efficient cooling techniques 
for that aim \cite{leibfried}.
For the typical frequencies and coupling constants, 
we will be focusing on the experimental 
implementation of Eq.~(\ref{hamrwa}) using $\rm Ca^{+}$ ions \cite{roos}. 
In these experiments, the electronic level separation is about THz 
while the trap frequencies are set typically in some MHz, 
and one order of magnitude smaller or higher by adjusting the trap potentials. 
For the classical Rabi frequency $\Omega$, a few MHz is also a realistic choice. 
We would also like to emphasize that our analysis and results are suitable to be applied 
to other known experimental setups such as those involving 
$\rm Be^{+}$ \cite{meekhof1996} 
or $\rm Yb^{+}$ \cite{olmschenk}.
The partition function in (\ref{partf1}) is a sum of an infinity number of terms which 
cannot be reduced analytically to a closed expression. 
Thus, a truncation is necessary. 
The convergence criterion for performing the truncation is 
explained in the note \cite{footnote2}. 
Each plot required a different number of terms kept in the sum, but in all cases the same
criterium is used.

\begin{figure}[!b] 
\includegraphics[width=8.5cm, trim = 10 10 0 0]{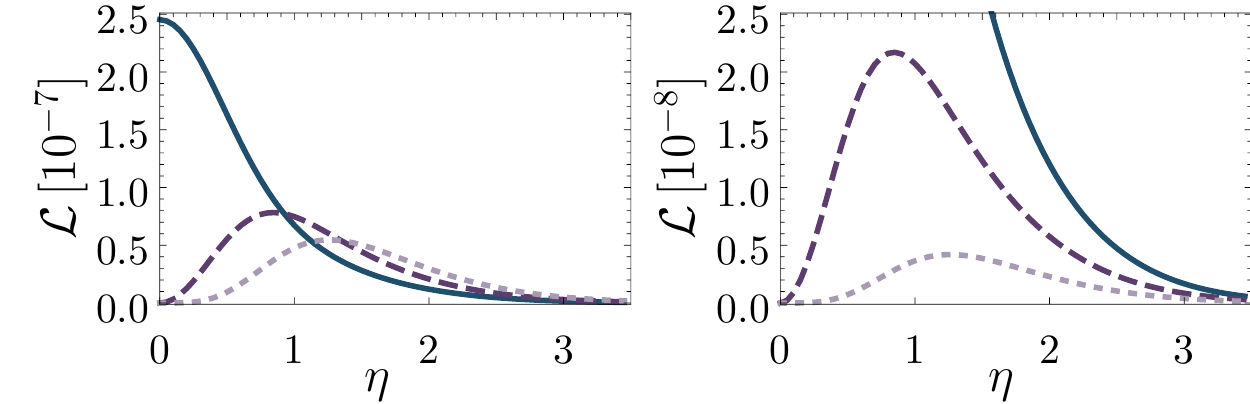} 
\caption{
(Color Online) NL as a function of 
the Lamb-Dicke parameter. The AJC (left) and JC (right) cases presented 
for $m = 0$ (solid), $ m = 1 $ (dashed) and $m = 2$ (dotted). The solid line 
is the same line in both panels. 
The used physical parameters were  
$\Omega =  1.0\pi {\rm MHz}$, 
$\omega_0 = 822.0 \pi {\rm THz}$, 
$\nu =5.0 {\rm KHz}$, $M=7.0\times 10^{-26}$kg, and $\bar n = 0.38$. 
Equation~(\ref{partf1}) is evaluated using the first 40 terms in the summation \cite{footnote2}. 
}                                                                                        \label{fig1L1} 
\end{figure}

The dependence of NL on the Lamb-Dicke parameter $\eta$  
is presented in Fig.~\ref{fig1L1} for a few values of $m$.  
The variation of $\eta$ in these plots comes from $\phi$ in Eq.~(\ref{etatrue}),  
since we are keeping the trap and laser frequencies fixed.
For small $\eta$, i.e., in the Lamb-Dicke regime, 
the Hamiltonians $\hat{\mathcal{H}}^{(m)}_{\pm} $ are basically 
ordinary Jaynes-Cummings models from cQED, in the sense that Eq.~(\ref{Omegaux1}) 
becomes approximately independent of the energy or number operator $\hat{a}^\dag\hat{a}$.  
In this regime, both the JC and AJC cases present the same ordering with respect to $m$. 
We see that the higher the sideband, or the number of motional quanta absorbed in 
the transition driven by the laser, the lesser the lag is. 
This means that   the sudden application of the laser becomes less irreversible 
and more like a quasistatic change. However, by increasing non-linearity, i.e., 
the magnitude of $\eta$, we depart from the ordinary cQED models, and Fig.~\ref{fig1L1} 
reveals that the JC and AJC present different responses with respect to irreversibility. 
For the JC case, increasing $\eta$ does not alter the order with respect to $m$ and, 
the higher the sideband, the lesser the NL. On the other hand, for the AJC such an 
order is not respected and, interesting enough, it comes to a point in which 
the higher the sideband, the higher the NL. Such behavior is induced by nonlinearity 
and it highlights well the different thermodynamic aspects resulting from JC and AJC 
models using trapped ions.
%

The behavior of the NL as $\eta$ is varied, with $\nu$, $\omega_0$ kept fixed, 
is determined by $f_n^m$ defined in Eq.~(\ref{auxf1}). In order to gain some insight 
about what was seen numerically in Fig.~\ref{fig1L1}, we now resort to analytical asymptotic limits.
For large $\eta$, the function $|f_n^m| \to 0$ because of the exponential 
in Eq.~(\ref{auxf1}). In this limit,  $\mathcal Z_{\pm}(\lambda_f) \to \mathcal Z(\lambda_i)$ so that 
$\mathcal L \to 0$.  

In the Lamb-Dicke regime, $\eta \ll 1$, 
we expand the exponential and Laguerre in Eq.~(\ref{auxf1}) up to second order in $\eta$ to find
\begin{eqnarray}                                                                         \label{fauxap}
\left|{f_n^m}\right|^{2} &\approx& \frac{(n+m)!}{n!m!^2} 
\left[1 - \frac{2n+m+1}{m+1} \eta^2 \right]{\eta^{2m}}.         
\end{eqnarray}
To obtain this expression we used $d L_n^m(x)/d x = - L_{n-1}^{m+1} (x)$ 
and $L_n^m(0) = (n+m)!/(n!m!)$ \cite{gradshteyn}. 
Now, by keeping just terms up to $\eta^2$ in $\left|f_n^m\right|^{2}$, one gets
\begin{equation}                                                                         \label{fauxap2}
\left|f_n^m\right|^{2} \approx [1 - (2n+1) \eta^2]  \delta_{m 0 } + 
(n+1) \eta^2 \delta_{m 1}.  
\end{equation}
Terms with  $m\ge 2$ appear only in higher powers of $\eta$. 
Notice that for  $m = 1$, $\left|f_n^1\right|^{2} \to 0$ and 
$\mathcal{Z}_{\pm}(\lambda_f) \to \mathcal{Z}(\lambda_i)$ as $\eta \to 0$, whhich makes 
$\mathcal {L} \to 0$.
On the other hand, $\left|f_n^0\right|^{2}$ in Eq.~(\ref{fauxap2})
is a concave function of $\eta$ with $\lim_{\eta\to 0}\left |f_n^0\right|^{2}=1$, $\forall n$.
Consequently, $\mathcal{L}\neq 0$ as $\eta\to 0$. All these features can be seen from  Fig.~\ref{fig1L1}.  
For $m > 1$, only higher order terms in $\eta$ contribute to $|f_n^m|$, forcing  
$\mathcal Z_{\pm}(\lambda_f) \to \mathcal Z(\lambda_i)$ as $\eta \to 0$, 
just like what happens when  $m=1$. 

The physical explanation for the distinct behavior found in the carrier transition  
$m=0$ lies in the system Hamiltonian after and before laser application. From Eq.~(\ref{Omegaux1}), 
one can see that 
\begin{equation}\label{etalim}
\lim_{\eta \to 0} \hat{\Omega}_{m}^\pm = \frac{\Omega}{2} \delta_{m 0}\id, 
\end{equation}
where $\id$ is the identity operator for the center-of-mass motion. 
By taking Eqs.~(\ref{hamct}) and (\ref{etalim}) into account, it follows that, 
when $m=0$, the laser is able to drive transitions between the two electronic states, 
even when $\eta = 0$. In other words, the pre- and post-quench Hamiltonians are different 
in the limit $\eta\to 0$, only when $m=0$. 
The process becomes then reversible in such a limit, provided $m\neq 0$.

\begin{figure}[!htbp] 
\includegraphics[width=8.5cm, trim = 10 10 0 0]{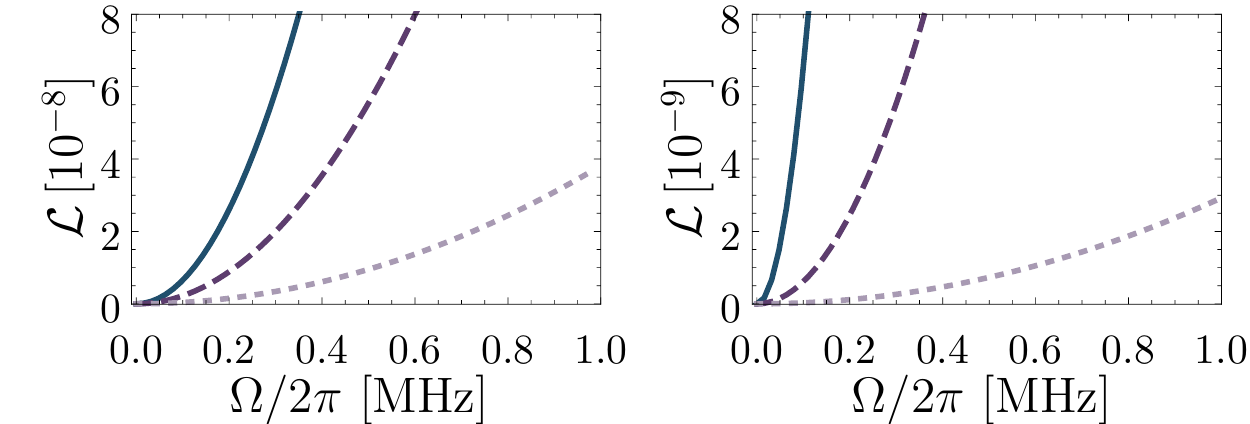} 
\caption{
(Color Online) Nonequilibrium Lag as a function of the 
Rabi Frequency of the AJC case (left) and JC case (right).
We use $\eta = 0.5$ and everything else is the same as in Fig.~\ref{fig1L1}. 
}                                                     \label{fig2L2} 
\end{figure}

Now, we investigate the role of the classical Rabi frequency 
 $\Omega$ on the irreversibility. 
The result is depicted in Fig.~\ref{fig2L2}, where one can see that the 
NL increases with $\Omega$. 
This behavior is expected from the detailed analysis of Eq.~(\ref{partf1}), and 
it can be physically understood from the fact that $\Omega$ is the work 
parameter and quantifies the intensity of the sudden quench.

\begin{figure}[!t] 
\includegraphics[width=8.5cm, trim = 10 10 0 0]{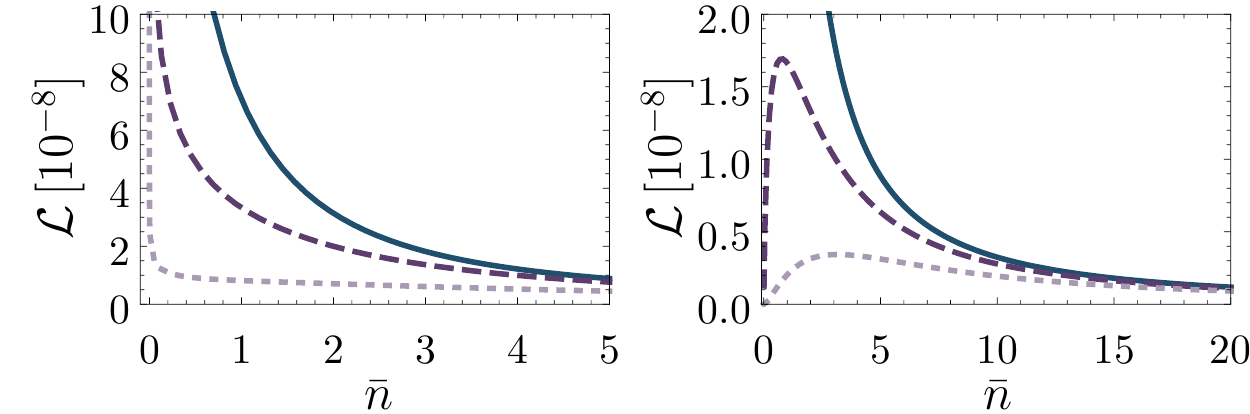} 
\caption{
(Color Online) Nonequilibrium Lag as a function of 
the mean occupation number of the initial thermal state 
for the AJC case (left) and for the JC case (right). 
We have used $\eta = 0.5$ and everything else is the same
as in Fig.~\ref{fig1L1}, except for  
the sum in Eq.~(\ref{partf1}) which is now evaluated with
$n = 2000$ for the plot on the left and 
$n = 5000$ for the one on the right \cite{footnote2}.
}                                                                                        \label{fig3L3} 
\end{figure}

In order to obtain a better understanding of the problem, it is necessary to go on and 
investigate the role of temperature. The NL as a function of the mean occupation number 
of the initial thermal state $\bar n$ in Eq.~(\ref{mocn}) is presented in Fig.~\ref{fig3L3}. 
It is noticeable that the AJC and JC models in the trapped ion system may respond so 
differently to variations of initial thermal energy of the system. 
In particular, it can be seen from Fig.~\ref{fig3L3} 
that the shown sidebands for the AJC and also for $m=0$ (which can be seen as either belonging 
to the AJC or JC classes) lead to a divergency in the NL as $\bar n\to 0$ ($\beta \to \infty$). 
This is not observed for for the JC case.   

Although the dependence of $\mathcal L$ on the temperature is a bit more intricate, 
since all factors in Eq.~(\ref{partf1}) depend on it, we again succeeded in providing an 
analytical treatment based on asymptotics that helps us to spot the reasons behind such 
different behavior found in the JC and AJC models. In the high temperature limit 
$\beta \to 0$ ($\bar n\to \infty$), a successive application of this limit, first 
to some exponentials and then to the hyperbolic functions in Eq.~(\ref{partf1}), results in
\begin{equation}                                                                         \label{limZ-T6}    
\!\!\!\frac{\mathcal Z_{\pm}(\lambda_f)}{\mathcal Z(\lambda_i)} \to 
\lim_{\beta\to 0} \left[ (\bar{n} + 1)^{-1} 
\sum_{n=0}^{\infty} \!{\rm e}^{-\beta \hbar \nu( n + \frac{m}{2})} \right]= 1, 
\end{equation}
which makes $\mathcal L \to 0$. This shows that, in this limit, the dynamics becomes reversible regardless of $m$.

For low temperatures, one can write 
$\cosh \beta x  \approx \tfrac{1}{2}{\rm e}^{\beta |x|}$ to find
\begin{equation}                                                                         \label{limZ-T3}    
\!\!\!\frac{\mathcal Z_{\pm}(\lambda_f)}{\mathcal Z(\lambda_i)} \!\to \!
\lim_{\beta\to\infty} \!\!
\left[\!(1\!-\!\delta_{m0}) {\rm e}^{-\tfrac{ \beta \hbar \omega_0 (1 \mp 1)}{2} } 
\!\!+\!\! \sum_{n = 0}^{\infty} {\rm e}^{- \frac{\beta \hbar}{2} \Phi_n^m} \!\right]\!, 
\end{equation}
where we have defined 
\begin{equation}                                                                         \label{funcphi}
\Phi_n^m := \nu(2 n + m) + \omega_0 - \sqrt{(\omega_{0} \! \mp \! m\nu)^{2} \! +\!
\Omega^{2}\left|f_n^m\right|^{2}} \, . 
\end{equation}
From this, we can analyze individually the AJC and JC cases. 
For the AJC and the carrier $m = 0$, 
it is easy to see that $\Phi_0^m < 0, \, \forall m$. 
As a result,
\begin{equation}                                                                         \label{limZ-T1}  
\lim_{\beta \to \infty}\frac{\mathcal Z_{-}(\lambda_f)}{\mathcal Z(\lambda_i)} = 
\infty, \,\,\, \forall m,    
\end{equation}
showing that for $\hat{\mathcal{H}}^{(m)}_{-}$ in Eq.~(\ref{hamrwa}) and 
$\hat{\mathcal{H}}^{(0)}$ in Eq.~(\ref{hamct}) the NL Eq.~(\ref{nlfinal}) always 
diverges when $\beta \to \infty$. For the JC case, we must give a closer look at the function 
$ \Phi_{n}^{m}$. From Eq.~(\ref{limZ-T3}), and
remembering that the case $m=0$ was already analyzed in Eq.~(\ref{limZ-T1}), 
\begin{equation}                                                                         \label{limZ-T}
\frac{\mathcal Z_{+}(\lambda_f)}{\mathcal Z(\lambda_i)} \to 
1 + 
\lim_{\beta\to\infty} \sum_{n = 0}^{\infty} {\rm e}^{- \frac{\beta \hbar}{2} \Phi_n^m}.  
\end{equation}
If, for a given $m$, at least one of the $\Phi_n^m$ appearing in Eq.~(\ref{limZ-T}) 
is negative, 
the above limit diverges and $\mathcal L \to \infty$.  
On the other hand, provided $\Phi_n^m \ge 0$ for all $n$, then 
\begin{equation}                                                                         \label{limZ-T2}
\frac{\mathcal Z_{+}(\lambda_f)}{\mathcal Z(\lambda_i)} \to k + 1,  
\end{equation}
where $k$ is the number of times $\Phi_n^m$ equals zero. 
Consequently, $\mathcal L \to {\rm ln}(1 + k)$ for the JC case.  
For the parameters chosen in Fig.~\ref{fig3L3},  
the JC case corresponds to $\Phi_n^m \ge 0$ 
and the limit in Eq.~(\ref{limZ-T2}) holds with $k = 0$, {i.e.}, 
no divergence is observed.  

Divergences of the NL can be understood, in general, as a consequence 
of the distinguishability between the post-work state and the reference thermal state 
used to evaluate the final free energy $F(\lambda_f, \beta)$. 
As previously commented, 
the NL can be written in terms of the relative entropy between those 
two states \cite{daffner}. 
As so, the smaller the NL, the more indistinguishable the two states are and, 
for orthogonal states, it diverges. 
In a quench process, as considered here, the initial state does not change 
after the work protocol \cite{fusco2014}. Consequently, the post-work state is a Gibbs state 
defined with inverse temperature $\beta$ and Hamiltonian (\ref{h0}). 
When $\beta\to \infty$, this state is basically $|0, g \rangle$. 
In the same limit, the reference thermal state used to evaluate $F(\lambda_f, \beta)$ 
will be given by the ground state of either the JC Hamiltonian or the AJC Hamiltonian, 
depending on the chosen $\omega_L$. For the physical parameters used in the simulations, 
the ground state of the JC Hamiltonian coincides with the post-work state which is 
$|0, g \rangle$, while  the ground state of the AJC Hamiltonian will be a 
superposition of $|0,e\rangle$ and $|m,g\rangle$. We can then see that NL will be smaller 
for the JC than for the AJC because the post-work state is more indistinguishable from 
the ground state of the former than from the ground state of the latter.  
All eigenstates and eigenvalues for AJC and JC Hamiltonians can be found in the Appendix.

\begin{figure}[!htb] 
\includegraphics[width=8.5cm, trim = 10 10 0 0]{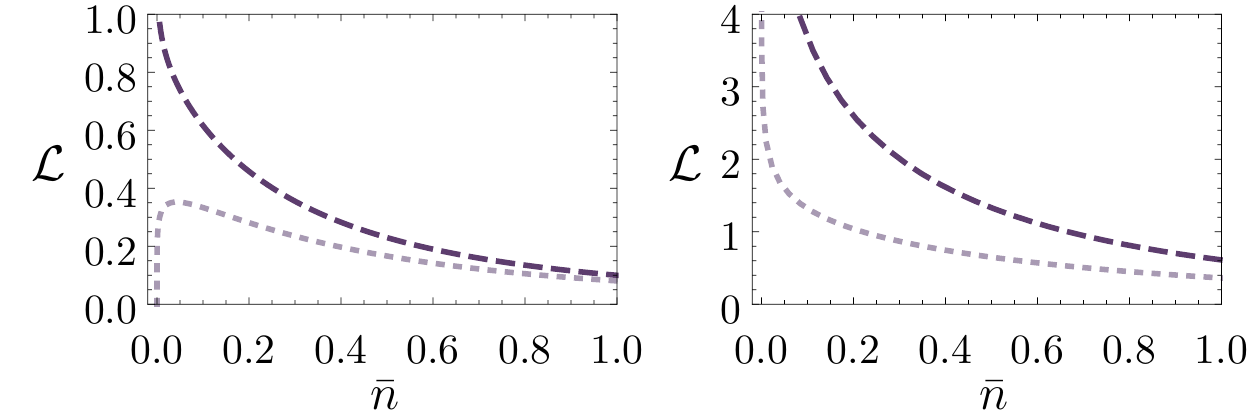} 
\caption{
(Color Online) Nonequilibrium Lag as a function of the mean
occupation number of the initial state 
for the JC case with $m = 1$ (dashed line) and for 
$ m = 2$ (dotted line). 
The parameters for the plot on the {\it left} are 
$\eta = 1.5$, 
$\Omega =  0.5 {\rm GHz}$, and
$\nu = 1.2 \omega_0 = 120 {\rm MHz}$, 
such that $\Phi_1^1 \le 0$ and $\Phi_n^2 \ge 0, \forall n$. 
For the plot on the {\it right}, the parameters are 
$\eta = 1.0$, 
$\Omega =  1 {\rm GHz}$, and 
$\nu = 1.2 \omega_0 = 120 {\rm MHz}$, 
such that $\Phi_1^1 \le 0$ and $\Phi_1^2 \le 0$. 
The sum in Eq.~(\ref{partf1}) is truncated in $n = 50$ 
terms \cite{footnote2}.
}                                                                                        \label{fig4L4} 
\end{figure}

We may wonder under which parameters choice 
the JC case can present divergences in the NL. 
In order words, how the system parameters can be chosen to cause at least one of the 
$\Phi_n^m$ in (\ref{limZ-T}) to be negative. 
The analysis of Eq.~(\ref{funcphi}) reveals that this is the case provided 
\begin{equation}                                                                         \label{limi1}
\left|f_n^m\right| > \frac{2}{\Omega} \sqrt{\nu (\omega_0+n\nu) (n+m)},  
\end{equation}
for a fixed $m$ (sideband) and some value of $n$. 
Now, in order to see this effect, one needs to go a bit beyond the current experimental 
set of parameters found in the literature. 
The result is shown in Fig.~\ref{fig4L4} where the parameters were deliberately chosen 
as to imply $\Phi_n^m < 0$ in some of the examples, 
making Eq.~(\ref{limZ-T2}) invalid and causing the NL to diverge  as $\beta \to \infty$.  
Although the parameters used to produce Fig.\ref{fig4L4} are unrealistic for the trapped 
ion system, one might think of their realization in an alternative system such as those 
in circuit quantum electrodynamics where ultrastrong couplings can be achieved. 
In this context, one may try to simulate the physics of trapped ions in the RWA using other 
controlled systems where such strong Rabi frequencies might be accessible.
%

\begin{figure}[!htbb] 
\includegraphics[width=6.0cm, trim = 0 15 0 0]{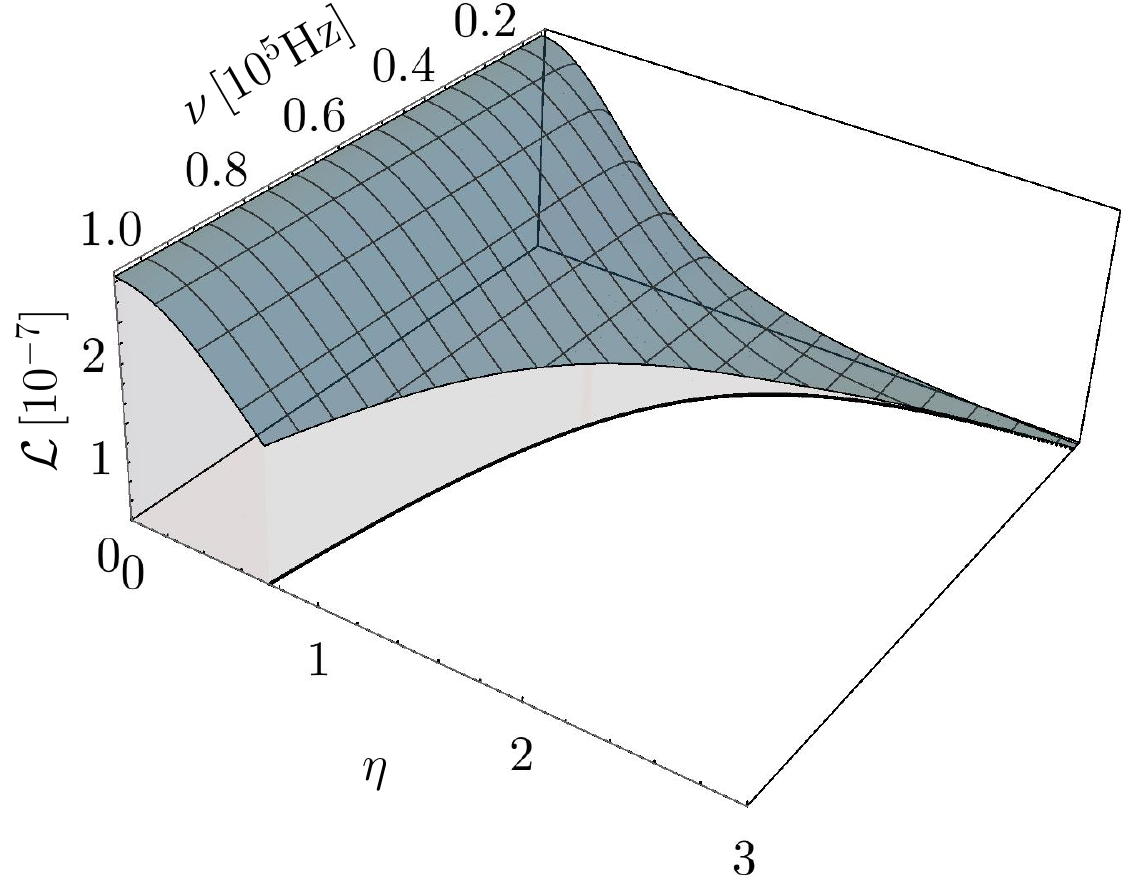} 
\caption{
(Color Online) Nonequilibrium Lag as a function of the natural 
frequency for the case $m = 0$. 
The solid plane curve is the projection of 
Eq.~(\ref{etatrue}) with $\phi = 0$ on the azimuthal 
plane $\mathcal L = 1.5 \times 10^{-7}$. 
The parameters for the plot are as in Fig.~\ref{fig1L1}.
}                                                                                        \label{fig5L5} 
\end{figure}

We now discuss the dependence of the Lamb-Dicke 
parameter on the trap frequency $\nu$ in Eq.~(\ref{etatrue}) and its implication for the 
irreversibility of the process. 
For that, we consider as one example the carrier transition $(m=0)$ in Fig.~\ref{fig5L5}
when, for a given trap frequency $\nu$, we vary $\eta$ from zero ($\phi = \pi/2$) 
to its maximum value ($\phi = 0)$. This is repeated for a broad range of trap 
frequencies. 
In general, the effect of varying the frequency of the trap is just to limit the maximum attainable 
values of $\eta$ obtained by changing the laser propagation direction in relation to the
trap axis (angle $\phi$). 
The NL basically does not change if $\nu$ is varied keeping $\eta$ fixed. 
Of course, according to Eq.~(\ref{etatrue}), in order to keep $\eta$ fixed while 
changing $\nu$, the angle $\phi$ must also be varied. 
For $\nu \ll \omega_0$ ($\nu \to 0$) one can adjust $\phi$ in order to keep $\eta$ 
constant. This limit, obtained from Eq.~(\ref{partf1}), reads
\begin{equation}                                                                         \label{limZ-nu}    
\!\!\!\!\frac{\mathcal Z_{\pm}(\lambda_f)}{\mathcal Z(\lambda_i)} \to 
{\rm sech} \tfrac{\hbar\beta\omega_0}{2}
\sum_{n=0}^{\infty} \cosh\!\!\left[ \tfrac{\beta \hbar}{2} 
     \sqrt{{\omega_0}^{2} \! +\! \Omega^{2}\left|f_n^m\right|^{2}} \right],                                                 
\end{equation}
regardless of being the JC or AJC case. Giving the convergence properties of $|f_n^m|$, 
discussed in \cite{footnote2}, this limit is finite. 
This finite behavior is illustrated with the case $m=0$ in Fig.~\ref{fig5L5}. 
Other choices of $m$ will lead to conclusions alike since the asymptotic behavior 
of $|f_n^m|$ with $n$ and $\eta$ does not depend on $m$ 
in any fundamental way \cite{footnote2}.
%

\begin{figure}[!htbp] 
\includegraphics[width=7.5cm, trim = 0 25 0 0]{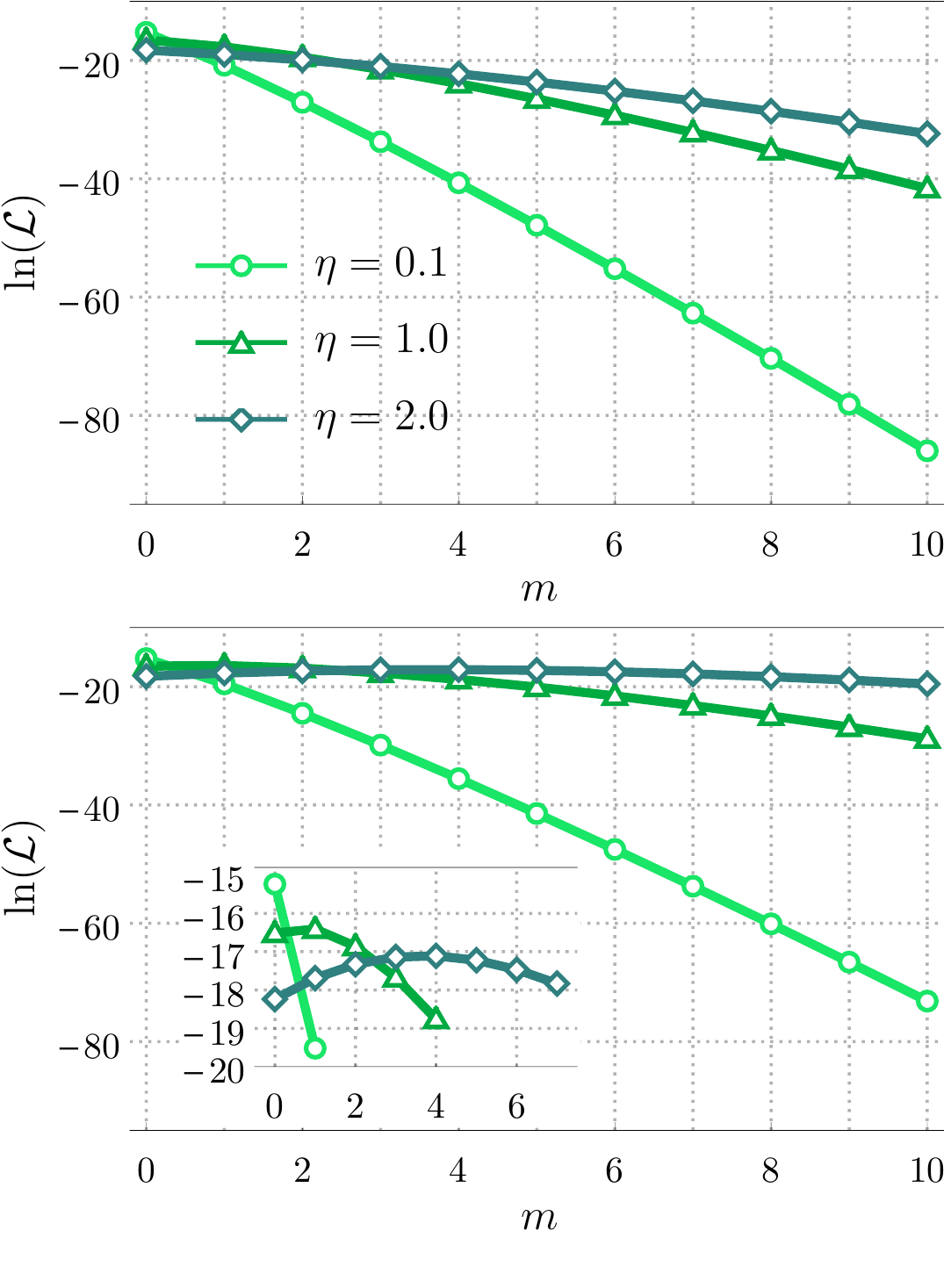} 
\caption{
(Color Online) Nonequilibrium Lag as a function of the 
sideband number $m$ for some illustrative choices of the Lamb-Dicke parameter.  
Top: JC case. Bottom: AJC case. 
The parameters are the same as in Fig.~\ref{fig1L1}. }                                   \label{fig6L6} 
\end{figure}

To finish the analysis of the NL, we explore its behavior for higher sidebands ($m > 2$). 
In Fig.~{\ref{fig6L6}}, we present a numerical study of such a dependence. 
One can see that the JC case tends to reversibility  
as the number of excitations $m$ exchanged between the ion motion and the electronic levels, 
induced by the laser, increases.  
For the AJC case, once again a rich behavior is found. For small $\eta$,  
the NL monotonically decreases with $m$, while for higher values of $\eta$, 
it comes to a point where the behavior is not monotonic anymore as highlighted 
in the inset of the bottom panel in Fig.~{\ref{fig6L6}}. 
From this point, we varied  $\eta$ up to $3.5$ (see Fig.~\ref{fig1L1}) to verify that, 
in this range, the maximum displaces to higher values of $m$ as $\eta$ increases. 
The same kind of analysis was performed considering the variation of $m$ 
for different temperatures and Rabi frequencies, and contrary to 
results in Fig.~\ref{fig6L6}, there are no remarkable differences between 
the AJC and JC cases. 

As a final remark, it is worthwhile to notice that, 
except for Fig.~\ref{fig4L4}, which is a theoretical 
extrapolation of the current experimental parameters, 
we have always found a higher NL for the AJC than for the JC. 
This can be once again understood from the relatively small values 
of $\bar{n}$ used in the simulations, and from the fact that the NL is
a relative entropy.  
This is the same reasoning we employed in the analysis of Fig.~\ref{fig3L3}. 
Additionally, the first order expansion of the hyperbolic 
function (in powers of $m\nu / \omega_0$) in Eq.~(\ref{partf1})
shows immediately that $\mathcal Z_- > \mathcal Z_+$. 
\section{Conclusions} \label{conc}                                    
From the point of view of  nonequilibrium thermodynamics, 
we studied the problem of sudden driving of a trapped ion by a classical laser field. 
This thermodynamical analysis was instrumental to pinpoint fundamental differences 
between the Jaynes-Cummings and Anti-Jaynes-Cummings-type Hamiltonians that arise 
in the trapped ion system by careful choice of the laser frequency. 
The role played by the magnitude of the Lamb-Dicke parameter, related to nonlinearity, 
and other physically relevant parameters was carefully studied. 
This makes our work useful also to the experimentalist who might be interested in the 
practical investigation of quantum thermodynamics of laser-manipulated trapped 
ion systems. In this respect, our work is, to the best of our knowledge, 
the first one to includes, in a thermodynamical approach, the great variety of possible 
electronic-vibration interactions available in the trapped ion system. 

Taking into account the small values of the NL encountered when using up-to-date
experimental parameters, noise in the experimental setup might impair its practical
determination. 
One way to circumvent that is to increase the Rabi frequency (intensity of the laser), 
since the NL increases monotonically with this parameter. 
To be more quantitative, a change of $\Omega$ from $10^6$ to $10^7$ is enough to increase 
the NL two orders of magnitude.

An experimental assessment of the findings of this paper might make use of a 2D trap 
(ion oscillations along $x$ and $y$ directions) and a driving laser coupling the 
electronic degrees of freedom to the $x$ motion. 
This can be easily achieved by choosing the right direction of the laser wave vector. 
The $y$ motion is used then as an ancilla in the interferometric scheme 
presented in \cite{various4}. For that, an extra laser is to be used to couple the system 
(electronic levels plus $x$ motion) to the ancilla in order to arrange for 
a proper gate entangling them \cite{various4}. 
With these, the work distribution can be experimentally determined and, with the help of 
the Jarzynski equality \cite{jarzinsky,crooks,tasaki}, 
the free energy and consequently the NL can be obtained.

\acknowledgments 
A.A.C. acknowledges to 
``Coordena\c{c}\~ao de Aperfei\c coa\-men\-to de Pessoal de N\'ivel Superior'' (CAPES).
FN, FLS and MP are supported by the CNPq ``Ci\^{e}ncia sem Fronteiras'' 
programme through the ``Pesquisador Visitante Especial'' initiative 
(Grant No. 401265/2012-9).
MP acknowledges financial support from John Templeton Foundation (grant ID 43467), 
the EU Collaborative Project TherMiQ (Grant Agreement No. 618074), and  also 
gratefully acknowledge support from  the COST Action MP1209 
``Thermodynamics in the quantum regime".
FLS is a member of the Brazilian National Institute of Science and Technology of 
Quantum Information (INCT-IQ) and acknowledges partial support from CNPq 
(Grant No. 307774/2014-7).
%
\section*{ Appendix } \appendix \label{ap}                            
\setcounter{equation}{0} \renewcommand{\theequation}{A-\arabic{equation}}   
In this appendix we  analytically perform the diagonalization of the Hamiltonians in 
Eq.~(\ref{hamrwa}) for any value of $m$. 
\subsection{Diagonalization of 
            \texorpdfstring{$\hat{\mathcal{H}}^{(m)}_{+}$}{}}  \label{do+}   
Let us consider the eigenbasis for the free Hamiltonian: 
$\{ | n , e \rangle, | n , g \rangle; n = 0,1,...,\infty  \} $. 
It is easy to see that the subspace spanned by the set 
$ \{ | n , e \rangle, | n + m, g \rangle \}$ is invariant under the action of 
the JC like Hamiltonian, $\hat{\mathcal{H}}^{(m)}_{+}$, in Eq.~(\ref{hamrwa}) 
 $\forall m, n$.  
Furthermore, if $m > n$ then it is true that  
\begin{equation}
\hat{\mathcal{H}}^{(m)}_{+} | n , g \rangle = 
\hat{\mathcal{H}}_{0} | n , g \rangle = 
\left( \hbar\nu n - \tfrac{\hbar\omega_0}{2} \right) | n , g \rangle, 
\end{equation}
{i.e.}, the eigenstate $| n , g \rangle$ must be included in the invariant subspace, 
which becomes   
$\{ | n , g \rangle, | n , e \rangle, | n + m, g \rangle \}$. 
Any matrix element of $\hat{\mathcal{H}}^{(m)}_{+}$ outside the 
invariant subspace is null because of (\ref{Omegme}). 
 
Taking the matrix elements of the Hamiltonian in the invariant subspaces,
and rearranging the basis, it acquires a simple block structure: 
\begin{equation}                                                                         \label{hblocks+0}
\hat{\mathcal{H}}^{(m)}_{+} = \hat{\mathcal{H}}_{+}^{[1]} \oplus 
                              \hat{\mathcal{H}}_{+}^{[2]}
\end{equation}
with 
\begin{equation}                                                                         \label{hblocks+}
\begin{aligned}
& \hat{\mathcal{H}}^{[1]}_{+} = 
\bigoplus_{n = 0}^{m - 1} \langle n , g |\hat{\mathcal{H}}^{(m)}_{+} | n , g \rangle 
= \hbar  
\bigoplus_{n = 0}^{m - 1} \left( \nu n - \tfrac{\omega_0}{2} \right), \\
& \hat{\mathcal{H}}^{[2]}_{+} \!=\! 
\bigoplus_{n = 0}^{\infty} \!\!\begin{pmatrix} 
                                          \!\! \langle n , e |\hat{\mathcal{H}}^{(m)}_{+} | n , e \rangle \!\!  & \!\! 
                                          \!\! \langle n\! + \! m, g |\hat{\mathcal{H}}^{(m)}_{+} | n , e \rangle \!\!\\
                                          \!\! \langle n , e |\hat{\mathcal{H}}^{(m)}_{+} | n\! + \! m , g \rangle \!\!  &  
                                          \!\! \langle n\! + \! m, g |\hat{\mathcal{H}}^{(m)}_{+} | n\! + \! m , g \rangle \!\!
                          \end{pmatrix} \\
& \,\,\,\,\, = \hbar
\bigoplus_{n = 0}^{\infty} 
                          \begin{pmatrix} 
                                          \nu n + \frac{\omega_{0}}{2}          & 
                                          \text{e}^{-i\omega_{L}t}\Omega f_n^m      \\
                                          \text{e}^{i\omega_{L}t} \Omega f_n^{m\ast} & 
                                          \nu (n+m)-\frac{\omega_{0}}{2}
                          \end{pmatrix},  
\end{aligned}
\end{equation}
with $f_n^m$ defined in Eq.~(\ref{auxf1}). 

The above block structure enables us to diagonalize the Hamiltonian by the 
diagonalization of each block. The first $m$ blocks in Eq.~(\ref{hblocks+}) 
are matrices of only one element having eigenvalues and eigenvectors, 
respectively, given by 
\begin{equation}                                                                        \label{eigval1+}
\zeta_{+}^{(n,m)} = \hbar\nu n-\frac{\hbar\omega_{0}}{2}, \,\,
\left|\zeta_{+}^{(n,m)}\right\rangle  = \left|n,g\right\rangle 
\end{equation} 
for each $n = 0,...,m-1$ for a given $m$.  
The following blocks in the diagonal block structure of (\ref{hblocks+0}) are 
$2 \times 2$ matrices, which can be diagonalized to give for all $n, m$ 
the eigenvalues 
\begin{equation}                                                                         \label{eigval+}
\begin{aligned}
\mu_{+}^{(n,m)} & = 
\hbar\nu\left(n+\frac{m}{2}\right)-
\frac{\hbar}{2}\sqrt{\omega_{L}^{2}+
\Omega^{2}\left|f_n^m\right|^{2}}, \\
\gamma_{+}^{(n,m)} & = \hbar\nu\left(n+\frac{m}{2}\right)+
\frac{\hbar}{2}\sqrt{\omega_{L}^{2}+\Omega^{2}\left|f_n^m\right|^{2}}, 
\end{aligned}
\end{equation} 
respectively, associated to the eigenvectors 
\begin{equation}                                                                         \label{eigevec+}   
\begin{aligned}
\left|\mu_{+}^{(n,m)}\right\rangle  & = 
\tfrac{\text{e}^{-i\omega_{\!L}t}\left[\omega_{\!L}-\sqrt{\omega_{\!L}^{2}+
\Omega^{2}\left|f_n^m\right|^{2}}\right]}{\Omega f_n^{m\ast}\,\sqrt{1+\frac{\left|\omega_{\!L}-
\sqrt{\omega_{\!L}^{2}+\Omega^{2}\left|f_n^m\right|^{2}}\right|^{2}}{\Omega^{2}\left|f_n^m\right|^{2}}}}
\left|n,e\right\rangle \\ 
& + \tfrac{1}{\sqrt{1+\frac{\left|\omega_{\!L}-\sqrt{\omega_{\!L}^{2}+
\Omega^{2}\left|f_n^m\right|^{2}}\right|^{2}}{\Omega^{2}\left|f_n^m\right|^{2}}}}\left|n+m,g\right\rangle, \\
\left|\gamma_{+}^{(n,m)}\right\rangle  & =\tfrac{\text{e}^{-i\omega_{\!L}t}\left[\omega_{\!L}
+\sqrt{\omega_{\!L}^{2}+\Omega^{2}\left|f_n^m\right|^{2}}\right]}
{\Omega f_n^{m\ast}\,\sqrt{1+\frac{\left|\omega_{\!L}+\sqrt{\omega_{\!L}^{2}+
\Omega^{2}\left|f_n^m\right|^{2}}\right|^{2}}{\Omega^{2}\left|f_n^m\right|^{2}}}}\left|n,e\right\rangle \\
& + \tfrac{1}{\sqrt{1+\frac{\left|\omega_{\!L}+\sqrt{\omega_{\!L}^{2}+\Omega^{2}
\left|f_n^m\right|^{2}}\right|^{2}}{\Omega^{2}\left|f_n^m\right|^{2}}}}\left|n+m,g\right\rangle ,
\end{aligned}
\end{equation}
where in this regime $\omega_{L} = (\omega_0 - m\nu)$.

\subsection{Diagonalization of  
            \texorpdfstring{$\hat{\mathcal{H}}^{(m)}_{-}$}{}} \label{do-}   
For the AJC like Hamiltonian, 
$\hat{\mathcal{H}}^{(m)}_{-}$, in Eq.~(\ref{hamrwa}), the invariant subspace is 
$\{ | n + m , e \rangle, | n , g \rangle \}$ for all $m,n$, while for $m > n$
it should be replaced by 
$\{ \left|n,e\right\rangle, | n + m , e \rangle, | n , g \rangle \}$.  %
Taking the matrix elements of the Hamiltonian in these subspaces, and rearranging the
basis as before, one finds
\begin{equation}                                                                         \label{hblocks-0}
\hat{\mathcal{H}}^{(m)}_{-} = \hat{\mathcal{H}}^{[1]}_{-} \oplus 
                              \hat{\mathcal{H}}^{[2]}_{-}, 
\end{equation}
where 
\begin{equation}                                                                         \label{hblocks-}
\begin{aligned}
& \hat{\mathcal{H}}^{[1]}_{-} = 
\bigoplus_{n = 0}^{m - 1} \langle n , e |\hat{\mathcal{H}}^{(m)}_{+} | n , e \rangle 
=  \hbar  
\bigoplus_{n = 0}^{m - 1} \left( \nu n + \tfrac{\omega_0}{2} \right), \\
& \hat{\mathcal{H}}^{[2]}_{-} \!=\! 
\bigoplus_{n = 0}^{\infty} \!\!\begin{pmatrix} 
\!\! \langle n , g |\hat{\mathcal{H}}^{(m)}_{+} | n , g \rangle \!\!  & \!\! 
\!\! \langle n\! + \! m, e |\hat{\mathcal{H}}^{(m)}_{+} | n , g \rangle \!\!\\
\!\! \langle n , g |\hat{\mathcal{H}}^{(m)}_{+} | n\! + \! m , e \rangle \!\!  &  
\!\! \langle n\! + \! m, e |\hat{\mathcal{H}}^{(m)}_{+} | n\! + \! m , e \rangle \!\!
                               \end{pmatrix}                                             \\
& \,\,\,\,\, = \hbar
\bigoplus_{n = 0}^{\infty} 
\begin{pmatrix} 
\nu (n+m) + \frac{\omega_{0}}{2} & 
\text{e}^{-i\omega_{L}t} \Omega f_n^m \\
\text{e}^{i\omega_{L}t} \Omega  f_n^{m\ast} & 
\nu n-\frac{\omega_{0}}{2}               
\end{pmatrix},  
\end{aligned}
\end{equation}
and $f_n^m$ is defined in Eq.~(\ref{auxf1}).                       

Now considering the one dimensional blocks where $m > n$, 
its eigenvalues and eigenvectors 
are, respectively, given by 
\begin{equation}                                                                        \label{eigval1-}                                                              
\zeta_{-}^{(n,m)} = \hbar\nu n+\frac{\hbar\omega_{0}}{2}, \,\,
\left|\zeta_{-}^{(n,m)}\right\rangle  = \left|n,e\right\rangle, 
\end{equation} 
for each $n = 0,...,m-1$ for a given $m$. 
The eigenvalues of each $2 \times 2$ blocks in Eq.~(\ref{hblocks-}) now becomes
\begin{equation}                                                                         \label{eigval-}
\begin{aligned}
\mu_{-}^{(n,m)} & =\hbar\nu\left(n+\frac{m}{2}\right)
-\frac{\hbar}{2}\sqrt{\omega_{L}^{2}+
\Omega^{2}\left|f_n^m\right|^{2}}  \\
\gamma_{-}^{(n,m)} & = \hbar\nu\left(n+\frac{m}{2}\right)+
\frac{\hbar}{2}\sqrt{\omega_{L}^{2}+
\Omega^{2}\left|f_n^m\right|^{2}}, 
\end{aligned}
\end{equation}
respectively, associated to the eigenvectors 
\begin{equation}                                                                         \label{eigevec-}
\begin{aligned}
\left|\mu_{-}^{(n,m)}\right\rangle  & = 
\tfrac{\text{e}^{-i\omega_{L}t}\left[\omega_{\!L}-
\sqrt{\omega_{\!L}^{2}+\Omega^{2}\left|f_n^m\right|^{2}}\right]}
{\Omega f_n^{m\ast}\,\sqrt{1+\frac{\left|\omega_{\!L}-
\sqrt{\omega_{\!L}^{2}+\Omega^{2}\left|f_n^m\right|^{2}}\right|^{2}}
{\Omega^{2}\left|f_n^m\right|^{2}}}}\left|n,g\right\rangle  \\
& +
\tfrac{1}{\sqrt{1+\frac{\left|\omega_{\!L}-
\sqrt{\omega_{\!L}^{2}+\Omega^{2}\left|f_n^m\right|^{2}}\right|^{2}}
{\Omega^{2}\left|f_n^m\right|^{2}}}}\left|n+m,e\right\rangle, \\
\left|\gamma_{-}^{(n,m)}\right\rangle & =
\tfrac{\text{e}^{-i\omega_{L}t}\left[\omega_{\!L}+
\sqrt{\omega_{\!L}^{2}+\Omega^{2}\left|f_n^m\right|^{2}}\right]}
{\Omega f_n^{m\ast}\,\sqrt{1+\frac{\left|\omega_{\!L}+
\sqrt{\omega_{\!L}^{2}+\Omega^{2}\left|f_n^m\right|^{2}}\right|^{2}}
{\Omega^{2}\left|f_n^m\right|^{2}}}}\left|n,g\right\rangle \\ 
& +
\tfrac{1}{\sqrt{1+\frac{\left|\omega_{\!L}+
\sqrt{\omega_{\!L}^{2}+\Omega^{2}\left|f_n^m\right|^{2}}\right|^{2}}
{\Omega^{2}\left|f_n^m\right|^{2}}}}\left|n+m,e\right\rangle ,
\end{aligned}
\end{equation}
for all $n, m$ and in this regime $\omega_{L} = (\omega_0 + m\nu)$.

As a final comment, the carrier transitions, Eq.(\ref{hamct}), 
eigenvalues can be obtained either from Eq.~(\ref{eigval+}) or from Eq.~(\ref{eigval-}), 
as its corresponding eigenvectors from Eq.~(\ref{eigevec+}) or Eq.~(\ref{eigevec-}) just setting $m = 0$. 



\begin{thebibliography}{99}
\bibitem{footnote2}
Numerical investigation of the sum in Eq.~(\ref{partf1}) shows that it 
is convergent since $|f^m_n|$, defined by Eq.~(\ref{auxf1}),
is a decreasing and oscillating function of $\eta$ and $n$. 
%
Some analytic progress is also possible to be made for large $n$ giving the well known 
asymptotic limit of Laguerre polynomials \cite{gradshteyn} appearing
in Eq.~(\ref{auxf1}). It can be shown that $|f^m_n| \sim n^{-1/4} \eta^{-1/2}$ 
in accordance with the tendency with $\eta$ and $n$ numerically revealed.
%
The sum is truncated as soon as the relative difference between 
successive terms 
become about $~10^{-10}$.
%
In this way, the number of terms kept in the sum may vary in 
different plots as it is clearly depends on the specific set of parameters used to produce the plot. 
We indicate the truncation number in the caption of each figure.
\end{thebibliography}
\end{document}